\documentclass[twocolumn,runningheads,natbib]{svjour2}
\journalname{Astrophysics and Space Science}

\smartqed  
\usepackage{graphicx}
\begin{document}

\title{Recent Progress on Anomalous X-ray Pulsars
}


\author{Victoria M. Kaspi}

\authorrunning{V. M. Kaspi} 

\institute{V. Kaspi \at
              Physics Department, McGill University, Montreal, Canada, H3A 2T8\\
              \email{vkaspi@physics.mcgill.ca}      
	      }

\date{Received: date / Accepted: date}

\maketitle

\begin{abstract}
I review recent observational progress on \\ Anomalous X-ray Pulsars,
with an emphasis on timing, variability, and spectra.  Highlighted
results include the recent timing and flux stabilization of
the notoriously unstable AXP 1E 1048.1$-$5937, the remarkable
glitches seen in two AXPs, the newly recognized variety of AXP variability
types, including outbursts, bursts, flares, and pulse profile changes,
as well as recent discoveries regarding AXP spectra, including their
surprising hard X-ray and far-infrared emission, as well as the pulsed radio
emission seen in one source.  Much has been
learned about these enigmatic objects over the past few years, with the pace
of discoveries remaining steady.
However additional work on both observational and theoretical fronts
is needed before we have a comprehensive understanding of AXPs and their
place in the zoo of manifestations of young neutron stars.
\end{abstract}

\section{Introduction}
\label{intro}

Although very few in number, the seven, and possibly nine, known
so-called ``Anomalous X-ray Pulsars'' (AXPs; see Table~\ref{ta:axps})
are potentially very powerful for making progress on the physics of
neutron stars.  AXPs embody properties that are highly reminiscent of
two other, very different classes of neutron star: the spectacular Soft
Gamma Repeaters (SGRs; see contributions by \linebreak S. Mereghetti and others in
this volume), and conventional radio pulsars.  The great similarity
of AXPs to SGRs is what makes the case for AXPs being magnetars
so compelling and also offers the hope of constraining magnetar physics. The
intriguing similarities with radio pulsars offer the promise of solving
long-standing problems in our theoretical understanding of the latter.

In this review, I describe the most recent observational progress on
AXPs.  The review will be divided into sections
on timing (\S\ref{sec:timing}), variability (\S\ref{sec:variability}),
and spectra (\S\ref{sec:spectra}), choices that, unfortunately, may
exhibit some personal bias, necessary given the limited space available.
I hope to show that recent AXP progress has been significant, however
ultimately much observational and theoretical work remains to be done
before a complete picture of AXPs and their place in the neutron-star
zoo becomes clear.

Note that the most comprehensive and recent review of AXPs and
magnetars in general is that by \citet{wt04}.  In this review, I make
use of the detailed, online summary of magnetar properties and references maintained at
McGill University by C. Tam\\ (www.physics.mcgill.ca/$\sim$pulsar/magnetar/main.html).

\begin{table*}[t]
\caption{Properties of Known and Candidate AXPs.}
\begin{center}
\label{ta:axps}
\begin{tabular}{llllllll}
\hline\noalign{\smallskip}
 Name  &  $P$   &  $\dot{P}^a$  & $B$  &  Timing$^b$ & X-ray Variability$^c$ & Waveband$^d$  & Association \\[3pt]
                & (s)  &($\times 10^{-11}$)& ($\times 10^{14}$~G)&  Properties  & Properties & Detections & \\
\tableheadseprule\noalign{\smallskip}
CXOU J010043.1$-$721134  & 8.02 & 1.9  & 3.9  & S?    & S       & X O? & SMC  \\
4U 0142+61             & 8.69 & 0.2  & 1.3  & S G?  & M P     & H X O I  & ... \\
1E 1048.1$-$5937       & 6.45 & 2.7  & 4.2  & N     & S F F B & H X I? & ... \\
CXOU J164710.2$-$455216 & 10.61 & ... & ... & ... & ... & X & Westerlund 1 \\
1RXS J170849.0$-$400910 & 11.00 &1.9  & 4.7  & S G G & S       & H X I & ... \\
XTE J1810$-$197        & 5.54 & 0.5  & 1.7  & N     & O B     & X I R & ... \\
1E 1841$-$045          & 11.78 & 4.2 & 7.1  & N      & S       & H X I? & SNR Kes 73 \\    
AX J1845$-$0258        & 6.97 & ...  & ...  & ...    & O       & X   & SNR G29.6+0.1  \\
1E 2259+586            & 6.98 & 0.048 & 0.59& S G    & S O B P & H X I & SNR CTB 109\\  
\noalign{\smallskip}\hline
\end{tabular}
\end{center}
$^a$Long-term average value.\\

$^b$S=stable (i.e. can be phase-connected over many-month intervals, generally); N=noisy (i.e. generally difficult to phase-connect over many-month intervals); G= one glitch. \\

$^c$S=stable (i.e. no variability, generally); M=modest variability; F=one flare; O=one outburst; B=bursts; P=pulse profile changes. \\

$^d$H=hard X-ray; X=X-ray; O=optical; I=infrared; R=radio. \\

\end{table*}

\section{Timing}
\label{sec:timing}

Timing observations of AXPs hold considerable information about both their
surroundings via the external torques they feel, as well as potentially about their
internal structure, via the ``glitches'' they experience.  Here we summarize what
is known regarding AXP timing properties, highlighting the most interesting issues.

\subsection{Stability and Phase-Coherent Timing}
\label{sec:stability}

Studies of the timing properties of AXPs can reasonably be categorized 
as pre- and post-Rossi X-ray Timing Explorer (RXTE), launched in
late 1995.  Prior to RXTE, timing studies were limited to occasional 
observations spread over many years.  These characterized the overall
spin-down behavior of several AXPs, and also suggested some
interesting deviations from simple spin-down \citep[e.g.][]{bs96,cm97,bss+00,pkdn00}. 
However the nature of these deviations could not be determined, because of
the paucity of observations.  Careful
searches for Doppler shifts of the observed periodicities had been
done \citep[e.g.][]{ikh92,bs96}; typical upper limits on $a\sin i$
were $\sim$0.1 lt-sec for a variety of orbital periods.

RXTE and its Proportional Counter Array (PCA) revolutionized
the timing of AXPs.  The first PCA studies of AXPs reduced the limits
on $a\sin i$ to 0.03 lt-s for 1E~2259+586 and 0.06 lt-s for 1E
1048.1$-$5937, effectively ruling out any main-sequence star companion
and rendering binary accretion models highly problematic
\citep{mis98}.  

Subsequently, a program of regular monthly monitoring of the AXPs with the PCA 
on RXTE was approved and showed that fully phase-coherent timing
of AXPs could be done over periods of years, assuming spin-down models consisting
of very few parameters \citep{kcs99}.  For example, \citet{gk02} showed that
in 4.5~yr of RXTE monitoring, pulse times of arrival for 1E~2259+586
could be predicted to within 1\% of the pulse period using $\nu$, $\dot{\nu}$ and
$\ddot{\nu}$ only.  Such stability is comparable to that of conventional young radio
pulsars and very much unlike the large amplitude torque noise commonly seen 
in accreting neutron stars.  Long-term, regular monthly (or even bi-monthly) observations
of AXPs continue today, with four of the five persistent confirmed Galactic sources
(4U 0142+61, 1RXS J170849.0$-$400910, 1E 1841$-$045 and 1E 2259+586) generally 
exhibiting stability that allows phase coherence with few parameters over years
\citep{klc00,gk02,ggk+02,kg03,dkg06}.  A summary of the timing properties of the known
AXPs is given in Table~\ref{ta:axps}.

\begin{figure}[h]
\centering
  \includegraphics[width=0.45\textwidth]{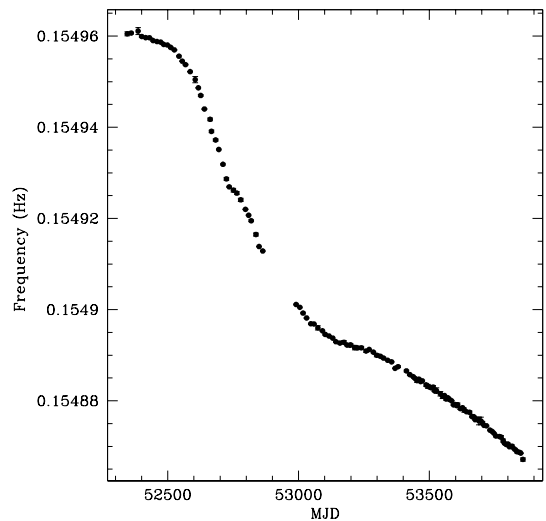}
\caption{Spin-frequency history of 1E 1048.1$-$5937 from
RXTE monitoring (work in preparation).  Note the
simple spin down in the last $1.9$~yr.}
\label{fig:1048freq}   
\end{figure}

\begin{figure}[h]
\centering
  \includegraphics[width=0.35\textwidth,angle=270]{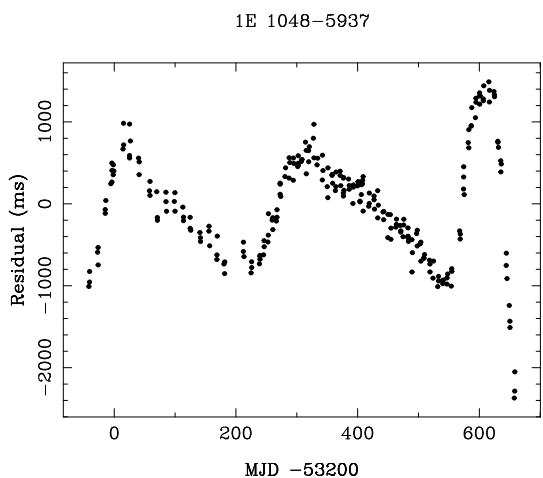}
\caption{Phase residuals following removal of four frequency derivatives 
for the last $1.9$~yr of data for AXP 1E 1048.1$-$5937.
The features are roughly 300~d apart and are presently consistent with 
being due to random timing noise (work in preparation).}
\label{fig:1048res}   
\end{figure}

One of the established persistent Galactic AXPs, 1E 1048.1$-$5937,
has been much less stable than the others, such that phase-coherent timing 
with unambiguous pulse
counting over more than a few weeks or months has been difficult 
to achieve \citep{kgc+01,gk04}.  This poor stability is apparent in
the source's frequency evolution (Fig. 1).  Recently, however,
during an extended period of pulsed flux stability following two
long-lived X-ray flares (see \S\ref{sec:flares} below), the timing
has also become quite stable, such that unambiguous phase coherence
could be maintained over a nearly 2-yr interval from MJD 53158 to
53858 using 4 spin parameters, although significant residuals remain
(Fig. \ref{fig:1048res}).  Details of these results will be described
elsewhere.  It remains to be seen if an end to this
stability will be accompanied by additional radiative events, a result
that would hopefully be useful for strongly constraining models for the
torque noise and flares.

\subsection{Glitches}
\label{sec:glitches}

\begin{figure}
\centering
  \includegraphics[width=0.4\textwidth]{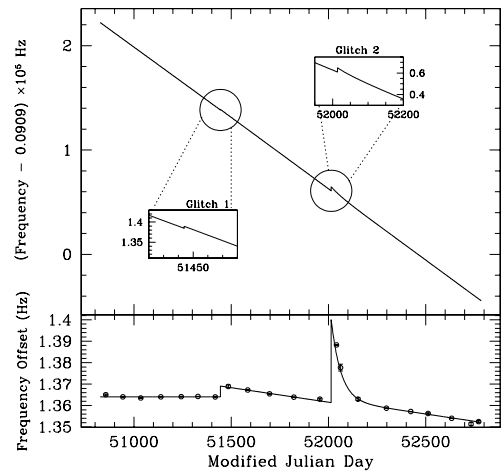}
\caption{Long-term RXTE timing data for 1RXS J170849.0$-$400910 showing
the two glitches observed thus far \citep[from][]{kg03}.}
\label{fig:1708_glitches}
\end{figure}

The impressive timing stability seen in most AXPs, in which pulse periods
could easily be determined to better than a part in a billion, 
permitted for the first time the unambiguous detection
of sudden spin-up glitches in AXPs.  1RXS J170849.0$-$400910 was the first
pulsar seen to glitch \citep{klc00}.   This glitch
had a fractional frequency increase of $6\times 10^{-7}$,
very similar to what is seen for the Vela pulsar and other
comparably young objects.  Continued monitoring of the same AXP revealed
a second, larger glitch 1.5~yr later \citep{kg03,dis+03}, with a nearly complete
recovery of the frequency jump, unusual by pulsar standards (see Fig. \ref{fig:1708_glitches}).

In June 2002, at the time of a major radiative outburst (see
\S\ref{sec:variability} below), the otherwise stable AXP \linebreak 1E 2259+586
exhibited a large glitch (fractional frequency increase of $4 \times 10^{-6}$)
which was accompanied by significant and truly remarkable recovery in which roughly
a quarter of the frequency jump relaxed on a
$\sim$40-day scale.  As part of that recovery, the measured spin-down rate of
the pulsar was temporarily larger than the long-term pre-burst average by
a factor of 2!  This recovery was somewhat similar to, though better sampled than, 
the second glitch seen in
1RXS J170849.0$-$400910.  The 1E~2259+586 event was the first (and still the only
unambiguous, but see \S\ref{sec:flares}) 
time a spin-up glitch in a neutron star was accompanied by any form of
radiative change.  This event suggests that large glitches in AXPs are
generally associated with radiative events; perhaps such an event occured
at the time of the second glitch in 1RXS J170849.0$-$400910
but went unnoticed due to sparse monitoring.
The very interesting recoveries of the 1E~2259+586 and second
1RXS J170849.1$-$400910 glitches seem likely to be telling us something
interesting about the interior of a magnetar and how it is different
from that of a conventional, low-field neutron star.  This seems worth
more attention than it has received in the literature thus far.

Recently a timing anomaly was reported in AXP \linebreak 4U~0142+61 \citep{dkgw06}.
Although at first the anomaly seemed consistent with a spin-up glitch
following a burst observed on April 6, 2006 \citep{kdg06}, the subsequent
data do not support a simple glitch interpretation.  This is work in progress and
will be reported on elsewhere.  However, \citet{mks05} and \citet{dkg06} suggest that
this pulsar may have glitched during a large observing gap in 1998-99, a possibility
supported by apparent changes in the pulse profile that seem
to have occured in the same interval \citep{dkg06}.

\section{AXP Variability}
\label{sec:variability}

Arguably one of the most interesting recent discoveries in the study of
AXPs is the range and diversity of their X-ray variability properties.
Pre-RXTE, there was variability reported \citep[e.g.][]{bs96,cm97,opmi98,pkdn00},
however those relatively sparse observations were made using different
instruments aboard different observatories, some imaging, some not, and
were often presented without uncertainties, making their interpretation difficult.
Moreover, given the sparseness of the observations, identifying a time scale
for the variations, or being certain the full dynamic range was being sampled,
was not possible.

Post-RXTE, and, additionally, with contemporaneous observations made
using Chandra and XMM, the picture has become much clearer.
Presently there appear to be at least four types of X-ray variability
in AXP pulsed and persistent emission:  outbursts (sudden large increases in the
pulsed and persistent flux, accompanied by bursts and other radiative and
timing anomalies, which decay on time scales of weeks or months), bursts (sudden
events, lasting milliseconds to seconds), long-term flux changes (time scales of
years), and pulse profile changes (on time scales of days to years).
We examine each of these phenomena in turn.

\subsection{Outbursts and Transients}
\label{sec:outbursts_transients}

\begin{figure}
\centering
  \includegraphics[width=0.5\textwidth]{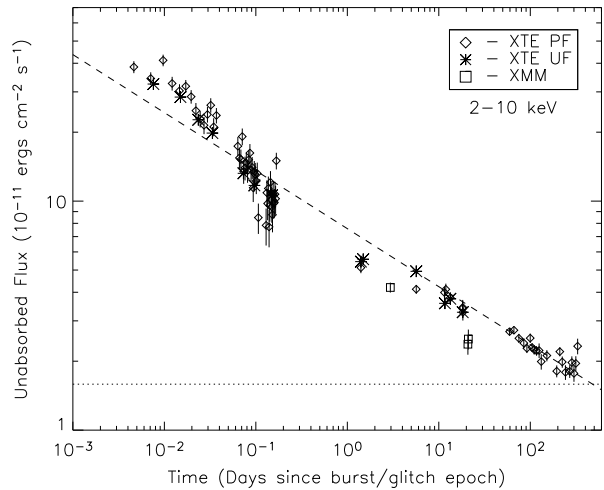}
\caption{Flux time series plotted logarithmically following the 2002 1E 2259+586
outburst \citep{wtk+04}.  Diamonds denote inferred unabsorbed
flux values calculated from RXTE PCA pulsed-flux measurements. Asterisks
and squares mark independent phase-averaged unabsorbed flux values from
RXTE and XMM, respectively. The dotted line denotes the flux level
measured using XMM 1 fortuitously week prior to the glitch. The dashed line
is a power-law fit to the PCA flux measurements during the observations
containing the burst activity ($<$1 day).}
\label{fig:2259_outburst}
\end{figure}

The current best example of an AXP outburst is that seen
in June 2002 from 1E~2259+586 \citep{kgw+03,wtk+04}.  This outburst,
which lasted only a few hours, fortuitously occured during a few-hour
monthly RXTE monitoring observation.  During the outburst,
the pulsed and persistent fluxes increased by a factor of $\sim$20 (see
Fig. \ref{fig:2259_outburst}), there were over 80 short SGR-like bursts
in a few-hour period (see \S\ref{sec:bursts}), there were substantial
pulse profile changes (see \S\ref{sec:profile_changes} below), there
was a short-term decrease in the pulsed fraction, the spectrum hardened
dramatically, there was a large glitch (see \S\ref{sec:glitches} above),
and there was an infrared enhancement (see \S\ref{sec:ir_variability}).  
All this came after over 4~yr of
otherwise uneventful behavior \citep{gk02}.  Note that had RXTE
not been observing the source during the outburst, the entire event
would have appeared, from the monitoring data, to consist principally
of a glitch.  With only monthly monitoring, all but the longest-term
radiative changes would have been missed.  Interestingly, this outburst
was notably different from SGR outbursts; for the AXP, the energy
in the outburst ``afterglow'' ($\sim 10^{41}$~erg; see Fig. \ref{fig:2259_outburst}) greatly
exceeded that in the bursts ($\sim 10^{37}$~erg).  This is in direct contrast to the giant
flares of SGRs.  The reason for this difference is unknown.

The outburst of 1E~2259+586 seems likely to be a good model for the
behavior of the ``transient'' AXP XTE J1810$-$197.  This source,
unknown prior to 2003, was discovered as a bright 5.5~s X-ray pulsar,
upon emerging from behind the Sun in that year \citep{ims+04}, and
has been fading ever since.  The source's spin down and spectrum are
consistent with it being an AXP.  \citet{ghbb04} showed from archival
ROSAT observations that in the past, the source, in quiescence,
was nearly two orders of magnitude fainter than in outburst.  See the
contribution by Gotthelf et al. in this volume for more details.
An important question raised by the discovery of XTE J1810$-$917 is
how many more such objects exist in our Galaxy?  This has important
implications for AXP birthrates.

Such an outburst may also explain the behavior of the unconfirmed
transient AXP AX J1845$-$0258 (see Table~\ref{ta:axps}).  Pulsations with
a period of 7~s were observed in an archival 1993 ASCA observation
\citep{tkk+98,gv98}, however subsequent observations of the target
revealed a large drop in flux, and pulsations have not been redetected.
Although no $\dot{\nu}$ has been measured for this source, it seems
likely to be an AXP given its period and location at the center of a supernova
remnant \citep{ggv99}.  It thus is plausible that the 1993 outburst was
similar to that of 1E~2259+586 or XTE J1810$-$197, although the likely dynamic
range for this source is unprecedented for an AXP.  See the contribution
to these proceedings by Tam et al. and \citet{tkgg06} for more details.

One of the most puzzling aspects of transient AXPs is why the
quiescent source is so faint.  In the standard magnetar model, the
requisite magnetic field decay energy input is persistent, as is the
magnetospheric twist for a fixed magnetar-strength magnetic field
\citep{td96a,tlk02}.  Although much attention has been paid to why
a magnetar's flux might skyrocket suddenly -- a sudden reconfiguration
of the surface following a crustal yield -- relatively little attention 
has been paid to how to stop or hide the bright X-ray emission from a 
neutron star presumably harboring a magnetar-strength field.

\subsection{Bursts}
\label{sec:bursts}

During the 2002 1E~2259+586 outburst, over 80 short, SGR-like X-ray
bursts were seen superimposed on the overall flux trend over the course
of the few-hour RXTE observation \citep{kgw+03}.  Some were \linebreak 
super-Eddington, though
only on very short (few ms) time scales.  In a detailed analysis of these
bursts, \citet{gkw04} found that in most respects, they are identical
to SGR bursts.  Specifically, the durations, differential fluence
distribution, the burst morphologies, the wait-time distribution,
and the correlation between fluence and duration, are all SGR-like.
However a few of the burst properties were different from those of SGR
bursts: for example, the AXP bursts had a wider range of durations, the
AXP bursts were on average less energetic than in SGRs, and the more
energetic AXP bursts have the hardest spectra -- the opposite of what
is seen for SGRs. Unlike in SGRs, the AXP bursts were correlated with
pulsed intensity. 

Bursts in AXPs were first reported by \citet{gkw02} who discovered two such
events in archival RXTE data from the direction of 1E~1048.1$-$5937.
A third burst, found nearly 3 yr later \citep{gkw06}, unambiguously
identified the AXP as the origin thanks to a simultaneous pulsed flux increase.
The bursting behavior in this source is not obviously correlated with any other property,
although we note tentatively that no bursting has been seen during the most recent
2~yr, when the pulsar has been exhibiting much improved timing 
(see \S\ref{sec:timing}, Fig.~\ref{fig:1048freq}) and also pulsed flux stability.
Perhaps the bursting was symptomatic of whatever activity also caused the
timing instability and pulsed flux flares (see \S\ref{sec:flares} below).

Four bursts have also been seen from the transient AXP XTE J1810$-$197
\citep{wkg+05}.  These events consisted of a $\sim$1~s spike followed by
an extended tail in which the pulsed flux was enhanced, similar to the
third burst of 1E 1048.1$-$5937 and a handful from 1E~2259+586.  
The XTE J1810$-$197 and 1E 1048.1$-$5937
bursts also showed a correlation with pulsed intensity, as did some of the 1E
2259+586 bursts.  

These observations led \citet{wkg+05} to hypothesize
that there are in fact two distinct classes of bursts, which they
named Type A and B.  Type A bursts are similar
to SGR bursts, in that they are uncorrelated with pulse phase and have
no extended tails.  Type B bursts, on the other hand, thus far observed
exclusively in AXPs, are correlated with pulsed intensity, generally
have long extended tails, and those tails tend to contain more energy
than the burst itself.  As both types of burst were seen in 1E~2259+586,
clearly they are not mutually exclusive, even during the same event.
\citet{wkg+05} speculate that Type A bursts have a magnetospheric trigger,
whereas Type B bursts originate from crust fractures.

Most recently, a total of five small, sub-Eddington X-ray bursts
have been seen between April and June 2006 from 4U 0142+61 \citep[][work in
preparation]{kdg06,dkgw06}.   The latter four, all within a single RXTE
orbit, were clearly accompanied by a pulsed flux increase (by a factor
of $\sim$4 relative to the long-term average) which decayed on
a time scale of minutes.  This suggests, as does the accompanying pulse
profile change and timing anomaly, that this source may have entered an
extended active phase.

With at least half of the known AXPs now established as capable of bursting, it
is clear that the production of occasional though clustered short SGR-like bursts is
a generic AXP phenomenon.

\subsection{Long-Term Flux Variations}
\label{sec:flares}

Variability in AXP 1E~1048.1$-$5937 had been reported for
years \citep[e.g.][]{cm97,opmi98,bss+00,mts+04}.  RXTE monitoring determined
the time scale of the changes and the morphology of
the pulsed light curve with far superior time sampling than in previous
studies \citep{gk04}.  Specifically, the RXTE monitoring
showed that over $\sim$7~yr, the source exhibited
two extended pulsed flux ``flares,'' the first lasting $\sim$100~days and the
second lasting over a year, each with rise times of several weeks.
Assuming a distance to the source of 5~kpc (which is not especially
well established), \citet{gk04} estimated the total energy released
in the pulsed components of the first and second flares to be 3 and 30
$\times10^{40}$~erg, respectively.  Subsequently, \citet{tmt+05}, using
XMM, which (unlike RXTE) is sensitive to pulsed fraction,
showed that in fact the pulsed fraction is anti-correlated with the
phase-averaged flux, suggesting the total energy released was at least
twice that in the pulsed component.

During these flares, the spin-down rate fluctuated by at least a factor
of 10 \citep{gk04}.  However, there was no obvious correlation between
the detailed evolution of the spin-down rate and flux.  Recently, while its flux
has been stable, 1E~1048.1$-$5937 has shown much greater timing stability
(see \S\ref{sec:timing} above).  This suggests that the large torque noise
and flux flaring were causally related; we must await another such event
to confirm this.

The flaring observed in 1E 1048.1$-$5937, a new phenomenon not yet
observed in any other AXP, is well understood in the context of the
twisted magnetosphere model \citep{tlk02}.  The flux enhancements can
be seen as being due to increased twisting of the magnetosphere by
currents originating from the stressed crust.  If so, a harder spectrum
is expected when the pulsar is brighter.  Unfortunately the existing
data cannot confirm this important prediction for this source.  
Decoupling of the torque from the pulsed flux can occur
in this model depending on the location of the enhanced magnetospheric
current; a current near the pole will have a disproportionate impact
on the spin-down.  \citet{gk04} argued that the absence of predicted
torque--luminosity variations in this source are problematic for models
in which the X-ray emission originates from accretion from a fossil disk
\citep[e.g.][but see \"U. Ertan et al.'s contribution]{chn00,alp01}.

The twisted magnetosphere model prediction that flux should be correlated
with hardness, though unconfirmed in 1E 1048.1$-$5937, does seem to be
borne out in observations of 1RXS J170849.0$-$400910 \citep[][see contribution
by N. Rea et al., this volume]{roz+05}.  Moreover, those authors suggest
that the epochs of greatest hardness occur near those in which glitches
were detected in this source (see \S\ref{sec:glitches}), with subsequent
softening post-glitch.  At least one additional glitch needs to occur, with
better observational coverage, before this conclusion is firm.

Recently, much longer-term radiative variations have been identified in
AXP 4U 0142+61, in which the pulsed flux appears to be increasing steadily
since 2000, such that there was a $\sim$20\% change by early 2006,
just prior to its exhibiting a sudden pulse profile change and bursts.
Interestingly, this variation does not seem consistent with magnetospheric
changes given the predictions of the twi- sted magnetosphere model.
This behavior is described in more detail in the contribution by R. Dib
et al. to this volume as well as in
\citet{dkg06}.  One particularly interesting implication of the increase
in X-ray flux from this source is that the phenomenon provides a simple
test of the irradiated fall-back disk model for the near-IR emission
(see \S\ref{sec:spectra_optir}).  If the source of irradiation, the AXP, increases in brightness,
the disk ought to as well.

\subsection{Pulse Profile Changes}
\label{sec:profile_changes}

The first observed pulse profile change in an AXP was reported by
\citet{ikh92} using {\it GINGA} data obtained in 1989.  They witnessed
a large change in the ratio of the amplitudes of the two peaks in
the pulse profile of 1E~2259+586, namely from near unity to over two.
They also reported a contemporaneous timing anomaly which, in hindsight,
is consistent with a spin-up glitch.

\begin{figure}
\centering
  \includegraphics[width=0.5\textwidth]{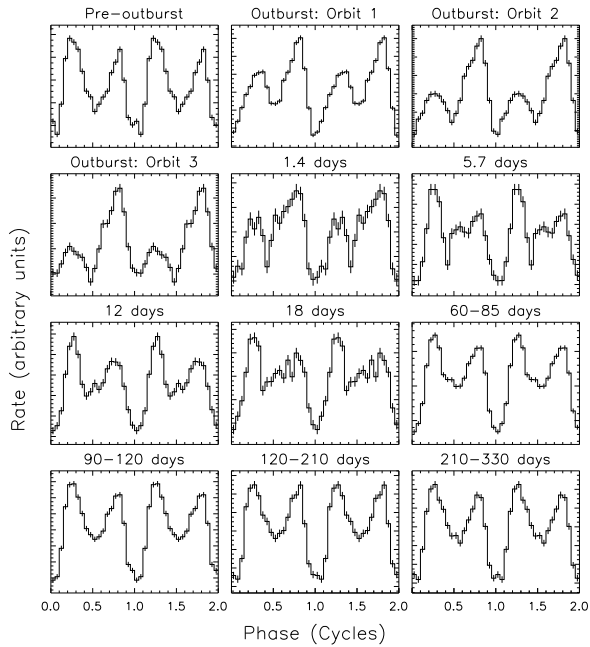}
\caption{Pulse profile changes in 1E 2259+586 following its 2002 outburst
\citep{wtk+04}.}
\label{fig:2259profile}
\end{figure}

A very similar pulse profile change was witnessed during and immediately
following the 2002 outburst of 1E~2259+586 \citep{kgw+03,wtk+04}.  Here,
the ratio of the amplitudes of the two pulses in the profile went from
unity pre-outburst to roughly two mid-outburst, relaxing back to normal
on a time scale of a few weeks (Fig. \ref{fig:2259profile}).  Curiously, 
the temporarily larger peak
in the 2002 outburst appeared to be the temporarily smaller peak in 1989, suggesting
that even if the physical origin of the events is the same, the details of the geometry
were different.  Given the nature of this event, a likely explanation for
the pulse profile change is a magnetospheric reconfiguration following
a crustal fracture that simultaneously affected the inside and
outside of the star.  This very strongly suggests that the \citet{ikh92}
pulse profile change was observed not long after a similar event; this is
consistent with their reported timing anomaly, and suggests such events
occur roughly every 1-2 decades in this source.

Most recently, long-term (i.e. time scale of several years) pulse
profile variations have been reported for AXP 4U~0142+61 \citep[][see
contribution by R. Dib et al, these proceedings]{dkg06}.  These accompany
a long-term pulsed flux increase (see \S\ref{sec:flares}).
The profile changes are consistent with a significant event having
occured some time between mid-1998 and 2000, in which the second and
higher harmonics of the profile increased dramatically, and have been
returning to their pre-event level ever since.  This gradual evolution
was, however, interrupted by an apparent sudden activity episode,
in which an SGR-like burst was detected along with a timing anomaly in
April 2006 \citep[see \S\ref{sec:bursts};][]{kdg06}.  The analysis of the
latter data are in progress.

\subsection{Near-IR Variability}
\label{sec:ir_variability}

The large number of recent near-IR detections of AXPs has revealed an
interesting new variability phenomenon in these sources.   Following the
2002 outburst of \linebreak 1E~2259+586, there was an infrared (K$_s$) enhancement
by a factor of $\sim$3, 3 days post-outburst.  This then decayed together
with the X-ray pulsed flux, with identical power-law indices \citep[see
Fig.~\ref{fig:2259_outburst}; \S\ref{sec:outbursts_transients}][]{tkvd04}.  Those authors
concluded that the origins of both flux increases were magnetospheric
\citep[but see contribution by \"U. Ertan, these proceedings] [for the
fossil-disk viewpoint]{ea06}.  A similar correlation was also reported
for AXP XTE~J1810$-$197 \citep{rti+04}.

Meanwhile, significant near-IR variability has also been reported for
AXP 1E~1048.1$-$5937 \citep{ics+02,wc02,dvk05}.  However, if anything, the
near-IR is {\it anti-correlated} with the X-ray flux.  This is puzzling
given the \linebreak 1E~2259+586 and XTE~J1810$-$197 results.  It is worth keeping
in mind, however, that there is now some evidence that 1E~1048.1$-$5937 has
been in an active phase from which it may have recently emerged (see \S\ref{sec:timing});
it will be interesting to see how its near-IR flux varies now that both
the X-ray flux and timing have stabilized (see Fig. \ref{fig:1048freq}).

In addition, \citet{dv06c} report on near-IR observations of 4U~0142+61
and report no apparent correlation with the X-ray pulsed flux.
They argue that the situation for this source
is unclear and warrants additional, more frequent observations.  Simultaneous 
far-IR observations
would also be of interest to establish conclusively that they originate from
a separate mechanism, namely radiation from a passive, irradiated fall-back
disk \citep{wck06}.

\section{AXP Spectra}
\label{sec:spectra}

The description of spectra of AXPs has, until very recently, been
limited to the soft (0.5--10~keV) X-ray band, as that is where
AXPs were discovered and have been traditionally studied.  However the 
recent discoveries of optical and near-IR emission have cast them
firmly into the multi-wavelength regime, and the even more recent
detections in the hard X-ray band as well as in the far-IR and
radio bands broadens the situation even further.

\subsection{X-ray Spectra}
\label{sec:xray_spectra}

In the traditional 0.5--10~keV X-ray band, AXPs have long been known
to show two-component spectra, which are well described by
a blackbody plus a power law \cite[e.g.][]{wae+96,ios+01,mskk03}.
It is currently thought that the blackbody arises from internal heating
due to the decaying intense magnetic field, while the non-thermal
component is a result of resonant scattering of the thermal
seed photons off magnetospheric currents in the twisted magnetosphere
\citep{tlk02}.  Detailed spectral modelling in this framework appears
to describe the spectrum of \linebreak 1E~1048.1$-$5937 \citep{lg06}, but
see the contribution by N. Rea in these proceedings.
Although some attempts were made to model the entire spectrum using
a single blackbody distorted by the effects of the intense magnetic
field on the atmosphere \citep[e.g.][]{oze01}, these could not
reproduce the non-thermal component adequately \citep[e.g.][]{phh+01,tlk02,lh03a}.

There is evidence supporting a physical connection between the two components,
such as the very slow evolution of the pulse profile with
energy \citep[e.g.][]{ios+01,gk02}.  For some AXPs (for example 1E~1048.1$-$5937
and 1E~1841$-$045), there is little if any variation in the pulse
profile, with no obvious difference between profiles in energy
bands that are thermally and non-thermally dominated.  Even 
in other AXPs for which the profile is energy-dependent, the
profiles in bands that are thermally and non-thermally dominated
are still very similar.  This is in stark contrast to the situation
for rotation-powered pulsars, for which the thermal and non-thermal components generally
have radically different X-ray pulse profiles \citep[e.g.][]{hsg+02a}.

Recently, \citet{hg05} have suggested on purely theoretical
grounds that the spectrum
of XTE J1810$-$197 is more appropriately described by a two-component
model consisting of two blackbodies (see also contribution by E. Gotthelf et
al. in this volume).  Their main argument for this
interpretation is that an extrapolation of the power-law component to
low energies greatly exceeds that expected if the seeds are thermal
photons from the surface, as the blackbody eventually runs out of
photons to supply.  Moreover, they argue, the expected blackbody cutoff
would then result in a substantial underestimate of the infrared flux,
assuming the latter is part of the non-thermal spectrum.  Very recently,
\citet{dv06b} have shown using independently measured interstellar column
densities that the intrinsic spectra really are cut off, i.e. the power-law
component does not in fact extend far below the observable band.  If correct,
the rationale for preferring the double blackbody over the blackbody plus
power law would not apply.

This section would not be complete without some discussion of the 
interesting recent results of \citet{dv06a,dv06b}, some of which are reported
on by Durant et al. in this volume.  Specifically, using high-resolution
X-ray spectroscopy, they identified absorption edges whose amplitudes
determine $N_H$ under reasonable assumptions, independent of the overall
continuum spectral modelling.  Using these newly measured values of $N_H$
and a novel distance estimation technique using reddening runs with distance
for red clump stars, they were able to first improve the spectral fits for 
several AXPs significantly, and second determine
much improved distance estimates for them.  Amazingly, among other things, 
they find that the
soft X-ray luminosities of practically {\it all} AXPs are consistent with the
value $\sim 1.3 \times 10^{35}$~erg~s$^{-1}$.  \citet{dv06a} argue that
this is consistent with the magnetar model's prediction
that there should be a saturation luminosity above which internal neutrino
cooling is at work.

\subsection{Hard X-ray Spectra}

A particularly interesting recent AXP discovery is that they are copious
hard X-ray emitters \citep{mcl+04,khm04}.  Though their spectra in 
$\nu F_{\nu}$ appear to fall off 
in the softer X-ray band, they turn up again above $\sim$10~keV.
The luminosities above 10~keV independently greatly exceed the available spin-down power
by factors of over 100.   This requires a new mechanism for accelerating particles
in the magnetosphere, in addition to an energy source, presumably the decaying
magnetic field.  SGR 1806$-$20 has also been detected in this energy range,
though interestingly it is somewhat softer than the AXPs \citep{mgmh05}.
\citet{khdc06} have further shown that the hard X-ray emission is a generic
property of AXPs, and for at least three sources, extends without a break up to
150~keV.  They also show from {\it COMPTEL} upper limits that the break must
lie under $\sim$750~keV.  Determining the location of the break could greatly
constrain emission models, possibly even providing independent evidence for
the magnetar-strength field.
 See the contribution by P. den Hartog et al. in this volume for details regarding
hard-X-ray emission from 4U~0142+61.

This hard X-ray emission, apart for being interesting for constraining the physics of 
magnetars, offers a unique way of detecting AXPs throughout the Galaxy,
since the soft X-ray emission suffers from high absorption, especially in the inner
Galaxy.  A focussing hard X-ray
instrument (like the NASA mission concept {\it NuSTAR})
would have the capability of detecting every magnetar in the Galaxy, provided their
hard X-ray emission is generic, even in quiescence.

\subsection{Optical and Infrared Spectra}
\label{sec:spectra_optir}

Currently, five of the known AXPs have been conclusively detected in the near-IR, with only
4U 0142+61 (the closest, least absorbed AXP) having been detected optically.
None of the SGRs has been detected optically, and only one has been seen
in the near-IR \citep[SGR 1806$-$20;][]{kot05}, and only during a particulary active phase.
For a summary of these detections, see \\
www.physics.mcgill.ca/$\sim$pulsar/magnetar/main.html and references therein.

The near-IR spectra of AXPs are an interesting puzzle.  First, given the
variability in the near-IR (\S\ref{sec:ir_variability}), piecing together an accurate instantaneous
spectrum using photometry requires contemporaneous observations, not always
possible.  Even more of a problem has been the generally unknown reddening
toward the sources, which have an enormous impact on the inferred intrinsic
spectrum. Nevertheless, some information regarding the optical and near-IR spectra
of AXPs is known. 
Overall, the major mystery is how the optical and near-IR emission connects with the X-ray
spectrum.  The blackbody emission seen in X-rays grossly underpredicts that in optical/near-IR,
while a simple extrapolation of the non-thermal component (when the spectrum is described
in this way -- see \S\ref{sec:xray_spectra}) generally overpredicts it.  
Expecting at least the optical emission to connect spectrally
with the X-rays is reasonable given that the latter is pulsed in 4U~0142+61 \citep{km02,dmh+05}
hence seems likely to originate in the magnetosphere, as does, presumably, the non-thermal X-ray emission.

Very recently, \citet{wck06} have shown using {\it Spitzer} data of 4U~0142+61 that in the far-IR, 
there appears
to be a component that is spectrally distinct from the near-IR emission.  They interpret this
component as resulting from a passive debris disk irradiated by the central X-ray source.  They suggest
that such disks, which originate from matter that falls back following the supernova explosion, 
may be ubiquitous around neutron stars.  They further suggest that the correlated near-IR/X-ray flux decay
observed by \citet{tkvd04} following the 2002 outburst of 1E~2259+586 and for XTE J1810$-$197 (see \S 3.5)
could be a result of
a disk around that AXP as well, a possibility also discussed by \citet{ea06} and \"U. Ertan et al. 
in 
these proceedings.  If the fallback-disk interpretation is correct for 4U 0142+61, the pulsed
X-ray flux change recently detected in this source (\S\ref{sec:flares}) may provide
an opportunity for a test of the disk hypothesis \citep[][and R. Dib et al. in this volume]{dkg06}, 
as there should be a correlated increase in the near-IR flux.  This idea requires that the overall 
X-ray flux, not just the pulsed component, also be increasing, which requires observations with an 
imaging X-ray telescope to verify.

\subsection{Radio Spectrum}

Very recently (in fact {\it after} this meeting took place!), \citet{crh+06} reported
the detection of radio pulsations from XTE~J1810$-$197.  This was a magnetar first and
in many ways a welcome discovery, having provided a nice radiative link between magnetars
and radio pulsars, in addition to the similarity already established from timing
observations (see \S 2).  It also demonstrated that pulsed radio emission can definitely be
produced in magnetar-strength fields in contrast to some predictions \citep[e.g.][]{bh98}.
Previous searches of other non-transient AXPs had come up empty \citep[e.g.][and see
contribution by M. Burgay et al., this volume]{mri+06,chk06}, suggesting the radio emission here
might somehow be related to this source's transient nature.  Given the small radio beaming
fraction reasonably expected for such slow pulsars, the absence of radio pulsations from
other sources could also be due to small-number statistics.

Also interesting is the unusual spectrum of the radio emission seen from XTE~J1810$-$197.
It has an unusually flat spectrum, with spectral index $>-0.5$, whereas radio pulsars have 
very steep spectra, with most indices between $-$1 and $-3$.   XTE~J1810$-$197 is the brightest radio 
pulsar known at frequencies $>$20~GHz.  Why this should be the case is an interesting new puzzle for
magnetar physics, one which has the potential to shed crucial new light on the long-standing
problem of the origin of radio emission in rotation-powered pulsars.

\subsection{Spectral Features}

\begin{figure}
\centering
  \includegraphics[width=0.3\textwidth]{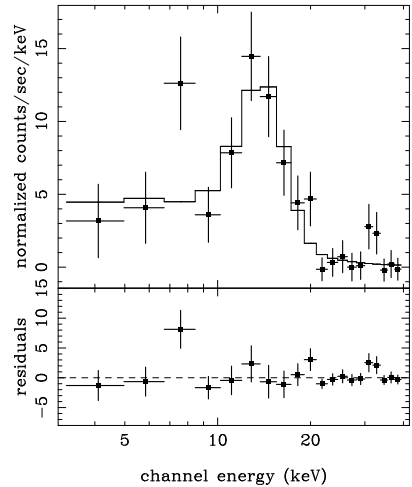}
\caption{Spectrum of the 1 s after the peak of the first burst observed from
1E 1048.1$-$5937 \citep{gkw02}.}
\label{fig:1048feature}
\end{figure}

Finally, there have been reports of features in AXP spectra.  The first such report
was for 1E~1048.1$-$5937 during the first of its two observed 2001 bursts \citep{gkw02}.  The
feature, which was most prominent in the first 1-s of the burst, was extremely
broad and seen apparently in emission at a central
energy of 14 keV (see Fig. \ref{fig:1048feature}). In terms of flux, it was comparable to the burst continuum
emission.  It was very significant; Monte Carlo simulations showed that such
a feature at any energy would be seen only 0.01\% of the time.  \citet{gkw06} saw a similar
feature in the third observed 1E~1048.1$-$5937 burst.
In that case, the central energy measured was also $\sim$13~keV.
In addition, \citet{wkg+05} observed a similar feature in one burst from XTE~J1810$-$197,
this time at energy 12.6~keV, with probability of chance occurence $4\times 10^{-6}$.
If interpreted as proton cyclotron lines, the energies of these features imply a magnetic
field above $\sim 10^{14}$~G, consistent with the magnetar hypothesis.  However
if the features are interpreted as electron cyclotron lines, the implied field is correspondingly $\sim$2000 times lower.
The latter would not necessarily be inconsistent with the magnetar interpretation, as it
is unclear what altitude above the neutron-star surface these lines originate.

Note that similar spectral features during SGR bursts have also been reported in the tails
of a few SGR bursts \citep[][see 
contribution by A. Ibrahim et al., these proceedings]{iss+02,isp03}.  However,
recently their statistical significance has been questioned (see contribution by
P. Woods et al.).

\citet{roz+05} reported spectral features at certain rotational phases from 
1RXS 170849.0$-$400910 from {\it BeppoSax} data.  These were claimed at the time
to be significant at the $\sim 4\sigma$ level.  
Observations with XMM of the same source saw no such features, implying either that
the {\it BeppoSax} features were spurious or that the effect is time variable.
See the contribution by N. Rea et al. in these proceedings for more information.

\section{Conclusions}
\label{sec:conclusions}

My hope in writing this review is to demonstrate that there remain many
important unsolved problems in the study of AXPs.  Overall, the basic issue of
what differentiates AXPs from SGRs remains, as does the origin of the intense
magnetic fields inferred.
Other important issues for which there simply was not enough space for
discussion here include the possible association of AXPs (and SGRs) with massive
star progenitors \citep[e.g.][see contribution by B. Gaensler et al.]{fng+05,mcc+06},
the puzzling lack of ``anomalously" bright X-ray emission from high-magnetic field radio
pulsars (see contribution by M. Gonzalez et al.), and the proposed connection
between magnetars and the so-called ``RRATS'' \citep[][and see contribution
by A. Lyne in this volume]{mll+06}.

As for how some of these problems will be solved, continued monitoring observations with
RXTE, as well as targetted studies with Chandra, XMM and INTEGRAL, will
obviously be of use.  Greater concerted efforts in the optical and near- and far-IR
are warranted, as are careful and repeated radio searches for transient pulsations
or other phenomena.  Finally, target-of-opportunity observations at the times of the
rare moments of AXP activity are definitely crucial for unravelling the overall physical
picture of these very interesting sources.

\begin{acknowledgements}
I thank C. Tam for maintaining the online McGill Magnetar Catalog
as well as R. Dib, F. Gavriil, and P. Woods for useful conversations.

\end{acknowledgements}

\bibliographystyle{apj2}
\bibliography{journals1,crossrefs,modrefs,psrrefs}   

\end{document}